# Epitaxial growth, magnetoresistance, and electronic band structure of GdSb magnetic semimetal films

Hadass S. Inbar*[1], Dai Q. Ho[2,3], Shouvik Chatterjee[4#], Mihir Pendharkar[4♦], Aaron N. Engel[1], Jason T. Dong[1], Shoaib Khalid[2†], Yu Hao Chang[1], Taozhi Guo[1‡], Alexei V. Fedorov[5], Donghui Lu[6], Makoto Hashimoto[6], Dan Read[4,7], Anderson Janotti[2], Christopher J. Palmstrøm[1,4*]

[1]Materials Department, University of California Santa Barbara, Santa Barbara, CA 93106, USA
[2]Department of Materials Science and Engineering, University of Delaware, Newark, DE 19716, USA
[3]Faculty of Natural Sciences, Quy Nhon University, Quy Nhon 55100, Vietnam
[4]Electrical and Computer Engineering Department, University of California Santa Barbara, Santa Barbara, CA 93106, USA
[5]Advanced Light Source, Lawrence Berkeley National Laboratory, Berkeley, CA 94720, USA
[6]Stanford Synchrotron Radiation Lightsource, SLAC National Accelerator Laboratory, CA, USA
[7]School of Physics and Astronomy, Cardiff University, Cardiff CF24 3AA, UK

(Dated: October 10, 2022)

Motivated by observations of extreme magnetoresistance (XMR) in bulk crystals of rare-earth monopnictide (RE-V) compounds and emerging applications in novel spintronic and plasmonic devices based on thin-film semimetals, we have investigated the electronic band structure and transport behavior of epitaxial GdSb thin films grown on III-V semiconductor surfaces. The $Gd^{3+}$ ion in GdSb has a high spin S=7/2 and no orbital angular momentum, serving as a model system for studying the effects of antiferromagnetic order and strong exchange coupling on the resulting Fermi surface and magnetotransport properties of RE-Vs. We present a surface and structural characterization study mapping the optimal synthesis window of thin epitaxial GdSb films grown on III-V lattice-matched buffer layers via molecular beam epitaxy. To determine the factors limiting XMR in RE-V thin films and provide a benchmark for band structure predictions of topological phases of RE-Vs, the electronic band structure of GdSb thin films is studied, comparing carrier densities extracted from magnetotransport, angle-resolved photoemission spectroscopy (ARPES), and density functional theory (DFT) calculations. ARPES shows hole-carrier rich topologically-trivial semi-metallic band structure close to complete electron-hole compensation, with quantum confinement effects in the thin films observed through the presence of quantum well states. DFT-predicted Fermi wavevectors are in excellent agreement with values obtained from quantum oscillations observed in magnetic field-dependent resistivity measurements. An electron-rich Hall coefficient is measured despite the higher hole carrier density, attributed to the higher electron Hall mobility. The carrier mobilities are limited by surface and interface scattering, resulting in lower magnetoresistance than that measured for bulk crystals.

## I. INTRODUCTION

In recent years, topological and trivial semimetals have received renewed interest due to observations of extremely large magnetoresistance (XMR) [1], which can be used in novel magnetic sensing technologies [2]. In topological semimetals, exotic transport characteristics due to the relativistic nature of charge carriers also hold great promise for applications in spintronic devices [3,4]. The coupling between antiferromagnetic order and relativistic quasiparticles offers new possibilities to control the symmetry of topological states and their spin-polarized currents by manipulating the magnetization orientation [5,6]. New spin-orbitronic devices based on current-induced Néel spin-orbit torques could dissipate less energy and perform at high switching rates [7]. Applications of antiferromagnetic topological crystals and magnetic topological heterostructures include spin valves [8,9], THz photodetectors [10–12], and recent studies are also exploring the design of heterogeneous catalysts leveraging the robust metallic surface states and spin polarization in electron transfer reactions [13–15].

Rare-earth monopnictides (RE-Vs) are a class of semimetals that can be easily integrated with III-V semiconductors [16–18] and present XMR [19], non-trivial topology [20], Fermi arcs [21], and unique magnetic phase diagrams [22] due to strong *p-f* and *d-f* electron coupling. The wide range of lattice constants, high thermodynamic stability, and the similarity of the rocksalt structure of RE-Vs with zincblende III-V semiconductors allow them to be incorporated epitaxially and processed into scalable devices [18,23,24]. III-V/RE-V magnetoresistive hybrid structures could utilize the geometric contribution to the extraordinary magnetoresistance found in high-mobility III-V-metal interfaces [25] and the tunability of conductivity in RE-Vs to optimize the heterostructure material parameters [26].

Many RE-Vs possess large non-saturating magnetoresistance, attributed to charge carrier compensation and the nearly-equal electron and hole high mobilities leading to a parabolic rise in magnetoresistance based on a classical two-band model [27,28]. To date, most thin-film reports of magnetoresistance values in RE-V [29,30] and other XMR semimetals [31] are significantly lower than their bulk crystal counterparts, potentially due to diminished mobilities as a result of surface and defect scattering, as well as possible deviations from exact mobility matching and carrier compensation. To explore the potential for novel RE-V semimetal devices, it is necessary to map the electronic structure of RE-V thin films and study the degrees of freedom that could help tune magnetoresistance, such as magnetic order transitions, quantum confinement effects, and defect scattering. From a thin-film synthesis perspective, GdSb is relatively straightforward to grow epitaxially on III-V semiconductors. Lighter rare-earth elements tend to be more reactive [32] and form more stable competing RE-$Sb_2$ phases [33].

Apart from magnetoresistive devices, synthesizing RE-Vs as thin films subject to biaxial strain and confinement effects also presents an opportunity to tune their band structure topology. While only RE-V bulk crystals with high spin-orbit coupling and large lattice parameters (XBi, X=La-Gd) were found to host topological

*Corresponding author. Email: hadass@ucsb.edu (H.S.I) cjpalm@ucsb.edu (C.J.P.) Present address: ♦ Department of Materials Science and Engineering, Stanford University, Stanford, CA, 94305 USA. # Department of Condensed Matter Physics and Materials Science, Tata Institute of Fundamental Research, Mumbai 400005, India. ‡Department of Physics, Princeton University, Princeton, New Jersey, 08540, USA. † Department of Physics, School of Natural Sciences, National University of Science and Technology, Islamabad 44000, Pakistan

semimetal states [34–36], recent studies suggest that Sb- and As-based RE-Vs subject to high pressure could also transition into a non-trivial topological phase [37,38]. Having a lattice parameter *a*=6.219 Å [39], GdSb is uniquely positioned between InSb (6.4794 Å) and GaSb (6.0959 Å)/AlSb (6.1355 Å), allowing tensile and compressive biaxial strain to be tuned by the underlying semiconducting III-V buffer layer structure. Further, epitaxial films of GdSb can serve as high-quality buffer layers that also aid as diffusion barriers for integrating reactive layers on III-V semiconductors [40]. The synthesis of thin films of GdSb also opens up more opportunities to study quantum size effects in RE-Vs, where quantum confinement was shown to alter carrier compensation and differentially affect the mobility of the electron and hole-like carriers [29,30,41]. In addition to improving the fundamental understanding of magnetotransport properties in GdSb, our electronic structure study can be used to engineer plasma resonance frequencies in RE-Vs and semimetal films for plasmonic mid-infrared optoelectronic applications [42].

Here, we report the growth, ARPES, and magnetotransport of epitaxial lattice-matched GdSb films grown on III-V buffer layers via molecular beam epitaxy (MBE). To our knowledge, this is the first report on the synthesis conditions and electronic properties of epitaxial GdSb thin films. As a member of the RE-V family, GdSb shares the common features of antiferromagnetic ordering [43] and an unusually high magnetoresistance [39], with previous magnetotransport reports of GdSb bulk crystals presenting magnetoresistance values up to 12500% [39]. Gd-V compounds serve as favorable model systems for studying magnetoresistive and magnetic scattering behavior in RE-V semimetals due to a relatively simple magnetic phase diagram, lack of orbital angular momentum, and deep-lying occupied 4f bands leading to a smaller *p-f* mixing than observed in Ce-Vs [22]. GdSb is a classical Heisenberg antiferromagnet (AFM), where $Gd^{3+}$ ions with S=7/2, L=0 order as a type-II AFM at 24 K [43,44], such that the Gd magnetic moments are ordered ferromagnetically along the <11-2> directions, and adjacent {111} planes are coupled antiferromagnetically (see **Figure 1(a)**, inset). The magnetic phase diagram of GdSb has an AFM phase that transitions to a spin-flop phase at very low fields (0.2 Tesla at 4K, B ∥ <001>). As the magnetic field increases, the spin-flop phase remains stable and linearly increases in magnetization until a critical field of 34.5 Tesla is reached [43].

**II. MBE GROWTH & EXPERIMENTAL METHODS**

The GdSb films were grown in a modified VG V80H III-V MBE growth chamber with a base pressure <5x10$^{-11}$ Torr. For magnetotransport measurements, epi-ready semi-insulating GaSb (001) wafers were used, with significant charge carrier freeze out expected below 80 K. For photoemission and scanning tunneling microscopy (STM) studies, conductive *p*-type Zn:GaSb (001) wafers were used, followed by *p*-type Be-doped III-V buffer layers. Surface reconstructions were monitored *in situ* with reflection high-energy electron diffraction (RHEED, **Figure S1**).

Coherent growth of lattice-matched films was studied *in situ* with RHEED and confirmed *ex situ* with x-ray diffraction (XRD) measurements. Streaky RHEED patterns were seen for all layers, suggesting smooth and epitaxial films. Surface cleanliness, stoichiometry, and Gd speciation were monitored with *in situ* X-ray photoelectron spectroscopy (XPS). XPS analysis of the Gd 4d, 3d, and Sb 3d core levels in **Figure 2(a-c)** confirms stoichiometry is achieved (Sb:Gd=1.04:1, within the uncertainty of XPS without a reference standard sample and accounting for the photoemission cross sections). The Gd 3d and 4d spectra and loss features are consistent with previous XPS studies of bulk GdSb [45]. The GdSb films did not show any signs of contamination *in vacuo*, i.e., the oxygen and carbon levels at the surface were below the XPS detection limit.

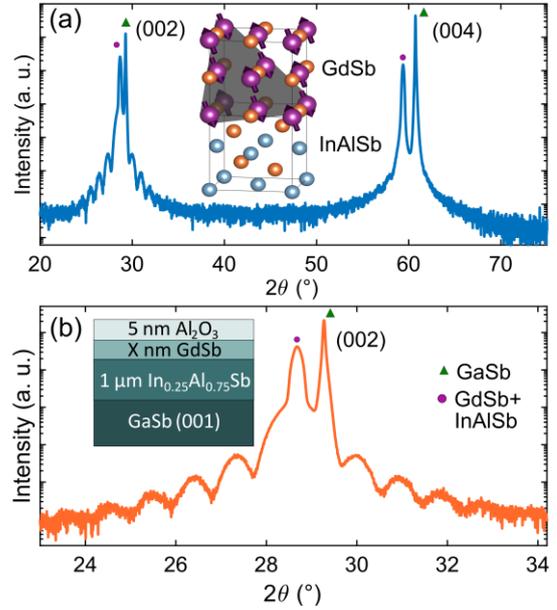

**Figure 1**. (a) Out-of-plane θ-2θ XRD scan for a 10 nm lattice-matched GdSb film grown on InAlSb/GaSb (001). The InAlSb and GdSb layers are indexed with a circle, the GaSb substrate with a triangle. Inset: Crystal structure and epitaxial relationship of the GdSb/III-V structure, with the magnetic structures for the type-II AFM ground state of GdSb shown. (b) Zoom in on the (002) reflection, inset shows the sample heterostructure.

Following growth and before removal from vacuum, the GdSb films were protected from degradation in the air by capping in another interconnected vacuum system with an amorphous $AlO_x$ layer deposited by e-beam evaporation of $Al_2O_3$ at room temperature. A schematic of a typical heterostructure grown for magnetotransport measurements is shown in **Figure 1**. More details on thin-film preparation, and the experimental growth window for GdSb are found in **Supp. Note 1**.

Vacuum ultraviolet (VUV) light ARPES measurements in the 20–100 eV range were performed on 4 nm and 20 nm thick- (A) *in vacuo* transferred GdSb films; and (B) Sb capped, air-exposed, and Sb decapped GdSb films. A custom-built vacuum suitcase with a base pressure <1x10$^{-10}$ Torr was used for transferring films from the growth chamber at UC, Santa Barbara, to beamline 10.0.1.2 at the Advanced Light Source (ALS) in Berkeley. At the SLAC National Accelerator Laboratory, Sb capped films were studied at beamline 5-2 at the Stanford Synchrotron Radiation Lightsource (SSRL). Ultraviolet photoemission spectra of the Gd 4f and Sb 4d core levels were collected for the *in vacuo* ALS transferred GdSb films and the SLAC

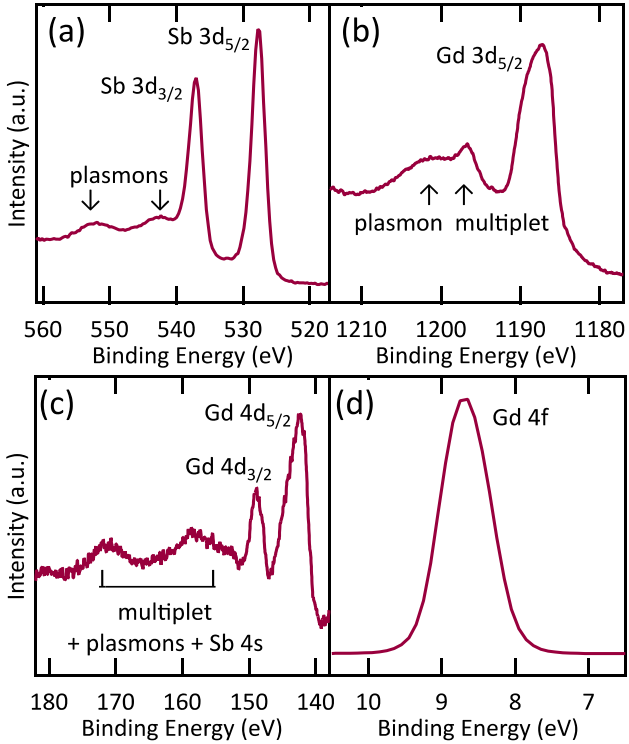

**Figure 2.** *In situ* XPS of the (a) Sb 3d, (b) Gd 3d5/2, (c) and Gd 4d core levels (d) UPS scan of the Gd 4f core level collected for *in-vacuo* transferred GdSb films.

Sb-capped films (after Sb desorption). We confirmed the chemical stability of the GdSb films in both cases by the absence of any oxidized components or oxygen 2s peak. In **Figure 2(d)**, a single Gd 4f peak is observed at 8.68 eV, indicating no oxidation.

We investigated the electronic structure of GdSb theoretically using density functional theory (DFT) and the screened hybrid functional of Heyd, Scuseria, and Ernzerhof (HSE06) [46,47] with 25% of exact exchange and accounting for spin-orbit coupling, as implemented in the VASP code [48,49]. The *4f* electrons were treated as valence electrons for the ferromagnetic (FM) and AFM calculations, whereas for the non-magnetic phase calculation, the *4f* electrons were treated as core electrons. We used the experimental lattice parameter of 6.219 Å for GdSb [39]. Additional details on the DFT calculations, ARPES data acquisition, and analysis are provided in **Supp. Note 2**.

### III. RESULTS AND DISCUSSION

#### A. Surface and structural characterization

**Figure 1(a)** shows a wide-range θ-2θ XRD scan of the GdSb film grown on a GaSb (001) substrate and $In_{0.25}Al_{0.75}Sb$ (referred to as InAlSb) buffer layer, revealing no additional peaks from impurity phases. A higher resolution triple-axis XRD scan near the (002) peak in **Figure 1(b)** shows exact out-of-plane lattice matching between the GdSb film and the InAlSb metamorphic buffer layer. The Pendellösung fringes indicate abrupt interfaces, and the extracted GdSb thickness values agree with *in situ* flux calibrations using RHEED oscillation.

The surface morphology and nucleation of the GdSb films were studied at room temperature with *in situ* scanning tunneling microscopy (STM) and *ex situ* atomic force microscopy (**Figure 3**). In a **Figure 3** crosshatch pattern originates from misfit dislocations in the underlying relaxed InAlSb buffer layer. The thin AlSb layer between the GdSb film and InAlSb buffer layer was added in an effort to expand the GdSb growth window to higher temperatures and lower Sb flux (**Figure S1(h)**) while mitigating any potential Gd-In interfacial exchange reactions [32] or displacement of In atoms to the surface [50]. **Figure 3(c-d)** shows a decrease in the surface roughness of 30 nm GdSb films grown with an AlSb interlayer, suggesting the higher stability of the AlSb surface at low $Sb_2$ overpressure could play a role in high-quality GdSb growth (see **Supp. Note 1.**). A relatively flat surface is achieved for the metamorphic buffer layer grown at low temperatures, as measured with *in situ* STM in **Figure 3(a)**). From the similar topography range in **Figure 3(b-d)**, a thickness of 5 nm of the amorphous $Al_2O_3$ capping layer appears to passivate well the GdSb surface without continuing reactions after removing the sample from a vacuum environment.

#### B. ARPES and DFT calculations

ARPES measurements performed for a 20 nm thick GdSb film are presented in **Figure 4** and **Figure 5**. In **Figure 4(b)**, we observe an ellipsoidal electron pocket (α) at the bulk X point, two nearly spherical light-hole (β) and spin-orbit split-off bands (γ) at the bulk Γ point, and a warped heavy-hole band (δ) resembling a square Fermi surface. The Fermi surface of the GdSb film is consistent with previous reports for other RE-Vs [27,35,36,51–53] and our DFT calculations in **Figure 4(g)**. Fits to all bulk bands near the Fermi level are presented in **Figure 5,** and the resulting Fermi wavevector values and calculated carrier densities (**Table 1**) are compared against DFT-extracted values (**Table 2**). The ARPES Fermi wavevectors are closer to the AFM phase DFT predictions than the non-magnetic phase. Based on the estimated Fermi volumes from the ARPES data, the 20 nm thick GdSb film shows similar hole and electron carrier densities, with an electron/hole ratio of $n_e/n_h$=0.84. The 4 nm thick GdSb films studied with ARPES showed a similar semimetallic band structure.

The 20 nm thick GdSb film displays multiple quantum well states in the hole bands (in **Figure 4(c-d)** and particularly visible near $\bar{\Gamma}$ for the γ pocket and $\bar{M}$ for the δ pocket in **Figure 5 (a-d)**) and electron pockets (**Figure 4(e)** and **Figure 5(e-h)**), confirming smooth conformal growth.

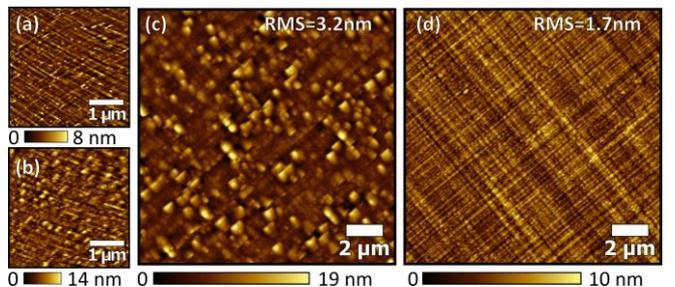

**Figure 3.** *In situ* STM images of (a) III-V lattice-matched buffer layer: 2 nm AlSb/200 nm $In_{0.25}Al_{0.75}Sb$/GaSb (001), and (b) 30 nm thick GdSb grown on the lattice-matched buffer layer structure. *Ex situ* atomic force microscope image of $Al_2O_3$ capped 30 nm GdSb films grown on the $In_{0.25}Al_{0.75}Sb$ layer (c) without and (d) with an AlSb interlayer.

At 20 nm, the GdSb film is still not at the bulk limit for all charge carriers, evidenced by the high number (>10) of finely-spaced quantum-well subbands in the δ hole pocket. The electron pocket shows fewer subbands crossing the Fermi level with a larger energy separation (**Figure 4(e)**) due to the smaller effective mass along the minor axis of the ellipsoidal electron pocket, suggesting that the electron band is strongly affected by quantum confinement.

Measurements of the electron pocket (**Figure 4(e)**) along $\bar{\Gamma} - \bar{M} - \bar{\Gamma}$ show both the expected $W - X_1 - W$ band dispersion from the $k_z = \Gamma$ zone center along the minor axis, as well as the neighboring Brillouin zone electron pocket band dispersions at $k_z = X_3$ along the electron pocket major axis $\Gamma - X_2 - \Gamma$. The nearly identical dispersions along the major axis of the electron pocket at different photon energies in **Figure 5(e-h)** result from the high $k_z$ broadening expected for the VUV light used in the ARPES measurements [36].

In **Figure 4(f),** the electron pocket effective mass enhancement along the major elliptical axis is calculated for the first four subbands near the Fermi level. Since thick GdSb films are expected to have weak electron correlation effects, the enhanced mass in the subbands with minima closer to the Fermi level is explained by the highly linear and non-parabolic dispersion of the electron pocket. A similar trend was recently observed for $IrO_2$ [54]. The ability to alter the total carrier density and average effective carrier mass in GdSb quantum wells through thickness-tuning suggests another route to controlling magnetoresistance in RE-V thin films [55].

The DFT HSE06 band structure calculations for the AFM and non-magnetic phases are presented in **Figure 4(g)** and appear to overlap at the Fermi level. The HSE06 AFM calculations matched our ARPES measurements and were shown to accurately describe the carrier concentrations and electronic bandgap at the bulk X point compared to other functionals [53,56]. See **Table 1** and **Table 2** for a comparison of ARPES and DFT Fermi wavevectors and band extrema at the X and Γ high symmetry points. The electron-hole band gap at the bulk X point (**Figure 5(e,f)**) is found to be $E_\alpha(X)-E_\delta(X)=0.21$ eV and is in close agreement with the AFM phase HSE06 calculations predicting a gap of 0.25 eV in **Figure 4(g)**. Near Γ, we see the most significant deviation between the AFM and non-magnetic band structure calculations. In the non-magnetic phase calculations, the spin-orbit coupled split-off $p_{1/2}$ (γ) pocket nearly crosses the Fermi level. However, as experimentally observed in **Figure 5(a-d)**, the γ pocket band maxima at the Γ point lies 0.19 eV below the Fermi level, showing that the predictions of the AFM phase (treating the p-f interactions explicitly) are in better agreement with experiments.

As a type-II AFM, the electron pockets in GdSb are not expected to show any exchange splitting because of the {111} orientation of the ferromagnetic planes. From the spectral weight for the AFM unfolded bands in **Figure 4(g)**, we can see the shadow band intensity along the $\bar{\Gamma} - L$ high symmetry axis is faint near the Fermi level. This can be explained by the relatively weak potential induced by the localized Gd *4f* spin structure and the itinerant Sb *5p* electrons forming the valence band [57]. In our ARPES data, we see no sign of band folding across the AFM Brillouin zone magnetic boundary (the $\Gamma - L$ axis projected onto $\bar{\Gamma} - \bar{X}$) in **Figure 4(c)** despite conducting the measurement below the Néel temperature (20 K), possibly due to either short-range AFM ordering or a low photoionization cross-section for the Gd *4f* level at 60 eV. Nevertheless, evidence of the strong *p-f* mixing predicted to take place in GdSb near the Fermi level [58] is found by comparing the position of the valence bands in DFT calculations for the non-magnetic phase versus the AFM phase. In the AFM phase, *p-f* mixing shifts the valence band downward and results in hole Fermi wavevectors which agree better with the experimental values.

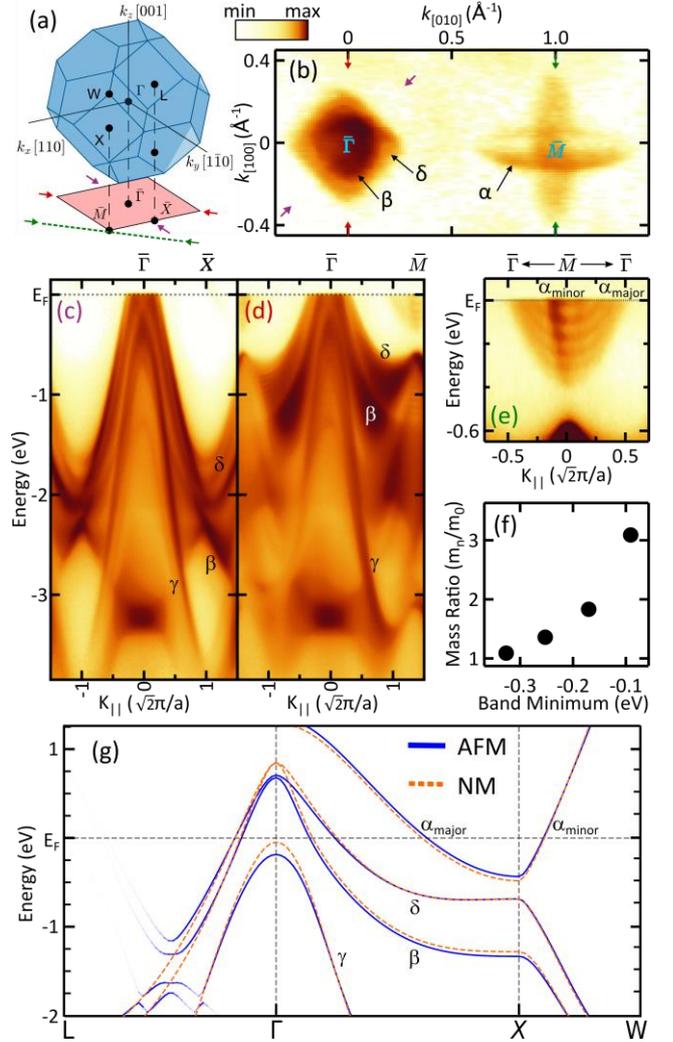

**Figure 4**. (a) Rocksalt bulk three-dimensional Brillouin zone and (001) surface projection, with E-k scan directions in panels (c-e) highlighted with arrows along $\bar{X} - \bar{\Gamma} - \bar{X}$ (purple, c) $\bar{M} - \bar{\Gamma} - \bar{M}$ (red, d) and $\bar{\Gamma} - \bar{M} - \bar{\Gamma}$ (green, e) (b) hv=60 eV ($k_z=\Gamma$) photoemission intensity plots at $E_F$ showing the Fermi surface map displaying hole-like bands (β, δ) and electron-like (α) bands. (c-d) Band dispersion of the hole pockets and (e) the electron pocket, presenting quantum well states in all three bands. (f) Relative mass enhancement of the electrons in the quantum well subbands ($m_n$) with respect to the lowest level mass ($m_0$) as a function of the band minimum energy. (g) DFT calculated band structure of GdSb in the antiferromagnetic (AFM), and non-magnetic (NM) states. The AFM gap is predicted to form along the <111> direction and the spectral weight of the AFM folded band along $\Gamma - L$ is reflected in the line transparency.

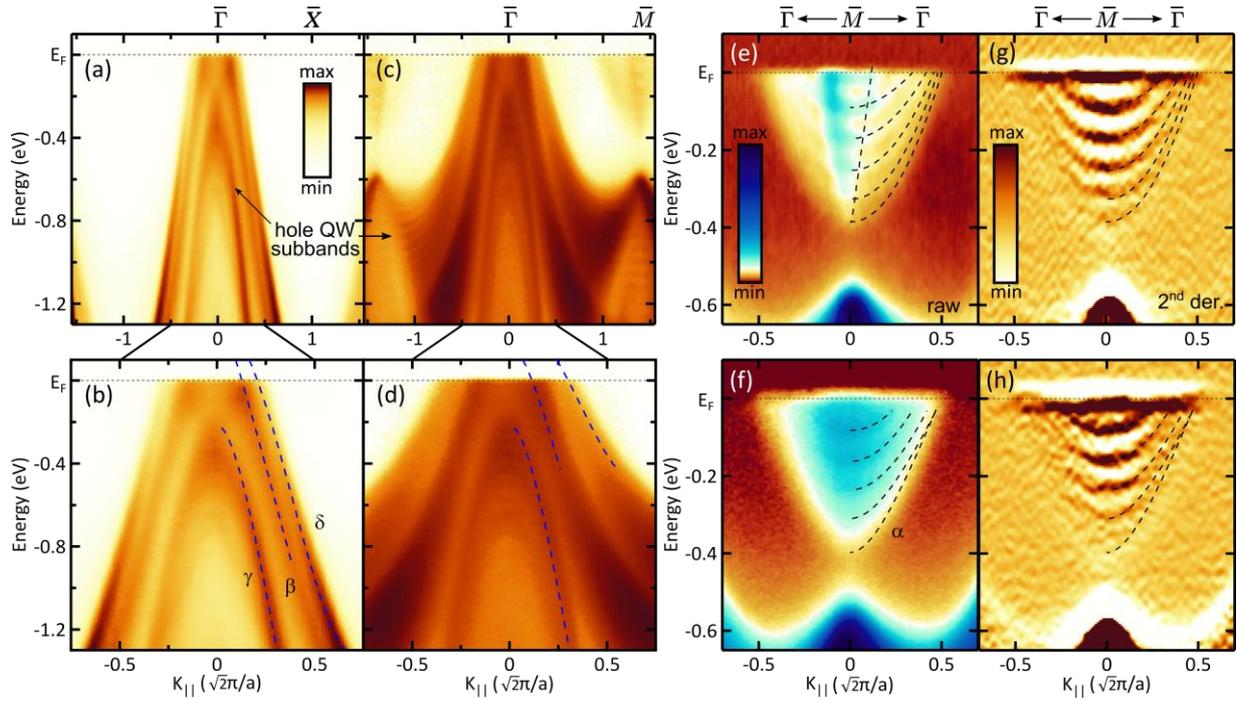

**Figure 5**. ARPES data and overlayed fit to the bands for positive wavevectors. The β, γ, and δ hole pockets along (a,b) $\overline{X} - \overline{\Gamma} - \overline{X}$, and (c,d) $\overline{M} - \overline{\Gamma} - \overline{M}$. The α electron pocket minor axis and major axis subbands along $\overline{\Gamma} - \overline{M} - \overline{\Gamma}$ at the (e) Γ plane ($h\nu$=60 eV) and (f) X plane ($h\nu$=94 eV). (g-h) 2nd derivative plots $\frac{\partial^2 I}{\partial E^2}$ of (e-f), respectively, enhancing the dispersive features of the raw data.

**Table 1:** Fermi surface of a 20 nm thick GdSb film. Band extrema energy positions, Fermi wavevectors, $k_f$, effective masses ($m^*$), and charge carrier densities (n) obtained from the ARPES measurements. Further details provided in the **Appendix** and **Supp. Note 2**.

| Fermi Surface | Band Extrema (eV) | | $k_F$ (Å$^{-1}$) | | $m^*$ ($m_0$) | | n ($10^{20}$cm$^{-3}$) |
|---|---|---|---|---|---|---|---|
| | Γ | X | | | | | |
| α | -- | -0.39 (±0.01) | Minor | 0.084 (±0.014) | Minor | 0.13(±0.03) | 3*$n_\alpha$ |
| | | | Major | 0.36 (±0.03) | Major | 1.73(±0.29) | 2.56 (±0.6) |
| γ | -0.21 (±0.01) | -3.12 (±0.01) | Does not cross the Fermi level | | | | |
| β | -- | -1.37 (±0.01) | $\overline{\Gamma} - \overline{M}$ | 0.10 (±0.01) | $\overline{\Gamma} - \overline{M}$ | 0.19 (±0.02) | 0.34(±0.04) |
| | | | $\overline{\Gamma} - \overline{X}$ | 0.100(±0.003) | $\overline{\Gamma} - \overline{X}$ | 0.17 (±0.01) | |
| δ | 0.31 (±0.06) | -0.60 (±0.01) | $\overline{\Gamma} - \overline{M}$ | 0.23 (±0.03) | $\overline{\Gamma} - \overline{M}$ | 0.40 (±0.08) | 2.72 (±0.8) |
| | | | $\overline{\Gamma} - \overline{X}$ | 0.170 (±0.002) | $\overline{\Gamma} - \overline{X}$ | 0.28 (±0.01) | |

**Table 2:** Band energies and Fermi wavevectors obtained from DFT HSE06 calculations for (a) the non-magnetic phase, (b) antiferromagnetic (AFM) phase, and (c) ferromagnetic (FM) phase (spin up/down bands accordingly). Carrier densities extracted from the DFT density of states of the non-magnetic phase are compared against the analytical calculation of carrier density using the ellipsoid model for the electron pocket, and the sphere model for the hole bands (marked with *). Further details provided in the **Appendix** and **Supp. Note 2**.

| Fermi Surface | Band Extrema (eV) | | $k_F$ (Å$^{-1}$) | | n ($10^{20}$cm$^{-3}$) |
|---|---|---|---|---|---|
| | Γ | X | | | |
| α | NA | (a) -0.43<br>(b) -0.43<br>(c) -0.13/ -0.61 | Minor, Major | (a) 0.103, 0.402<br>(b) 0.101, 0.383<br>(c) 0.054/0.111, 0.205/0.461 | (a) 4.20 (4.32*)<br><br>(c) 3.61 |
| γ | (a) -0.05 (b) -0.19<br>(c) -0.13/-0.26 | (a) -3.31<br>(b) -3.30<br>(c) -3.15/-3.33 | | | |
| β | (a) 0.85<br>(b) 0.67<br>(c) 0.74/0.44 | (a) -1.28<br>(b) -1.33<br>(c) -1.29 | $\overline{M} - \overline{\Gamma}$ | (a) 0.151 (b) 0.139 (c) 0.140, 0.133 | (a) 1.18 (1.08*)<br><br>(c) 0.92 |
| | | | $\overline{X} - \overline{\Gamma}$ | (a) 0.144 (b) 0.139 (c) 0.148/0.125 | |
| δ | (a) 0.85<br>(b) 0.67<br>(c) 1.04/0.91 | (a) -0.68<br>(b) -0.68<br>(c) -0.63 | $\overline{M} - \overline{\Gamma}$ | (a) 0.254 (b) 0.244 (c) 0.255/0.255 | (a) 3.01 (3.62*)<br><br>(c) 2.68 |
| | | | $\overline{X} - \overline{\Gamma}$ | (a) 0.187 (b)0.183 (c) 0.193/0.178 | |

Finally, DFT calculations of a forced ferromagnetic phase in GdSb (**Figure S4**), which forms under high magnetic fields, predict significant exchange splitting in the electron *d* band almost leading to a *p-d* band crossing which would result in a non-trivial band topology and the emergence of Weyl points. Due to the low position of the potential band crossing with respect to the Fermi level (~0.5eV), any Weyl physics contribution to transport is expected to be negligible.

## C. Magnetotransport

The temperature dependence of the electrical resistivity for a 10 nm thick GdSb (001) film is shown in **Figure 6(a)**. The total resistivity of the stack peaks at 135 K as the InAlSb buffer layer charge carriers freeze out, indicating the GdSb film has become the lowest resistive path for transport. Upon further cooling, the resistivity adheres to the same trends observed for bulk RE-V crystals [59]: a linear decrease with temperature is observed down to the Néel temperature, where a kink in resistivity at T=25 K ($T_N$=25.02 and 24.55 K for 10 and 4 nm thick films, respectively) is followed by a sharper decrease in resistivity. The kink and the sharp drop in resistivity below the Néel temperature indicate that spin-disorder scattering significantly contributes to the total resistivity close to the Néel temperature. Applying higher magnetic fields perpendicular to the film plane increases the resistivity, mainly at lower temperatures.

The small Hall resistivity in **Figure 6(d)** and high longitudinal resistivity upon applied magnetic field (**Figure 6(b)**) both suggest that the films are stoichiometric and have a low defect concentration [39]. Macroscopic scattering sites in LuSb thin films have been shown to promote *p*-type Hall behavior [60], and non-stoichiometric Gd-Vs produce strong negative magnetoresistance behavior at low fields ($\mu_0$H<1T) as well as a smaller magnetoresistance [39].

The magnetoresistance at 14 T reaches a maximum value of 110% for a 10 nm thick film at 1.8 K (**Figure 6(b)**) and 10% for a 4 nm thick film (**Figure S7**), the latter being the highest value yet reported for RE-Vs with a thickness <5 nm [29,30]. Shubnikov-de Haas (SdH) oscillations were observed on top of the magnetoresistance background and were obtained by removing the background using a 5th-order polynomial fit (**Figure S5**). The fast Fourier transform (FFT) of the quantum oscillations reveals three frequencies in **Figure 6(c)**, corresponding to the circular cross-section of the ellipsoid electron pocket at the X bulk point along the film plane normal $\alpha_\perp$ (359 T) and the two hole pockets, namely, β (740 T) and δ (1564 T) at the Γ point. The Fermi surface of the electron pockets lying at the film (001) plane is not resolved in the FFT spectrum due to the high effective mass of the major-axis band (see **Table 1**). Therefore, we have used the DFT calculated aspect ratio of the elliptical electron pocket to estimate its Fermi volume. The FFT amplitude corresponding to the electron pocket $\alpha_\perp$ shows a single peak, despite significant exchange splitting. Further information on the SdH oscillations, magnetotransport models, as well as details on the absence of exchange spin splitting signatures in the electron pocket FFT frequencies are provided in the **Appendix** and **Supp. Note 3**.

Carrier concentrations derived from SdH oscillations are provided in **Table 3**. A similar disparity between ARPES and magnetotransport calculations was seen in earlier RE-Vs studies [28,51,52]. The lower Fermi wavevectors extracted from ARPES fits are explained by significant $k_z$ broadening at the VUV wavelengths, leading to an underestimated Fermi surface area [36]. The carrier concentration calculated from both SdH oscillations and the magnetotransport multicarrier fit are in good agreement with previous values of ~4.2 $10^{20}$cm$^{-3}$ found for GdSb bulk-crystals [39]. A carrier compensation of $n_e/n_h$=0.67 is calculated from the SdH oscillations, similar to the ARPES-extracted ratios, yet still far from the nearly exact charge compensation typically assumed for bulk RE-V crystals. Deviations from an exact carrier ratio $n_e/n_h$=1 were measured in other bulk RE-V compounds such as DySb [61], LaBi [62], YSb [52], and NdSb [63]. Only a moderate level of compensation may be in fact needed for RE-V compounds to exhibit XMR as long as the carrier mobilities are high. Additional causes for uncompensated carrier concentrations could be more systematic, as the calculations of carrier ratios from Fermi surfaces could depend on the Fermi volume estimation [27]. Lastly, the density of states values from the DFT calculation in **Table S1** show nearly exact carrier compensation. Yet, the computed carrier densities in GdSb also depend significantly on the Fermi level position. A 50 meV shift would cause a relatively steep change in $n_e/n_h$ from 0.7 to 1.5. RE-V thin films are more susceptible to quantum confinement effects and Fermi level shifts than bulk crystals due to potential charge transfer at the surface/interface.

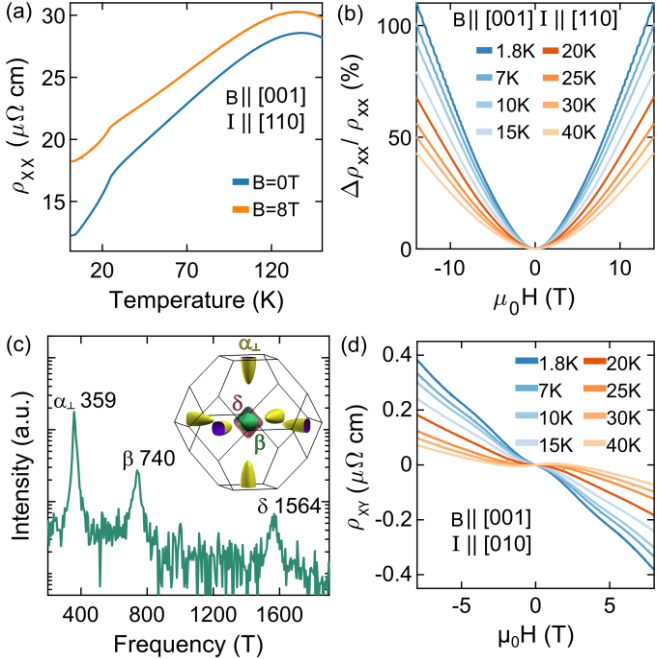

**Figure 6**. Magnetotransport behavior in a 10 nm thick GdSb film, (a) Temperature dependence of the longitudinal resistivity in the epitaxial stack (b) Magnetoresistance vs. magnetic field at temperatures below (blue) and near/above (orange) the Néel temperature. (c) Log plot of the FFT spectrum of SdH oscillations measured at 1.8 K. Inset shows the GdSb Fermi surface. (d) Temperature dependence of the Hall resistivity.

**Table 3:** Fermi surface parameters extracted from SdH oscillations for a 10 nm thick GdSb film, calculations of the values are described in **Supp. Note 3**.

| Fermi Surface | Frequency [Tesla] | $k_F$ (Å$^{-1}$) | n ($10^{20}$cm$^{-3}$) |
|---|---|---|---|
| $\alpha_\perp$ minor axis | 359 | 0.104 | 4.32 |
| β | 740 | 0.150 | 1.14 |
| δ | 1564 | 0.251 | 5.34 |

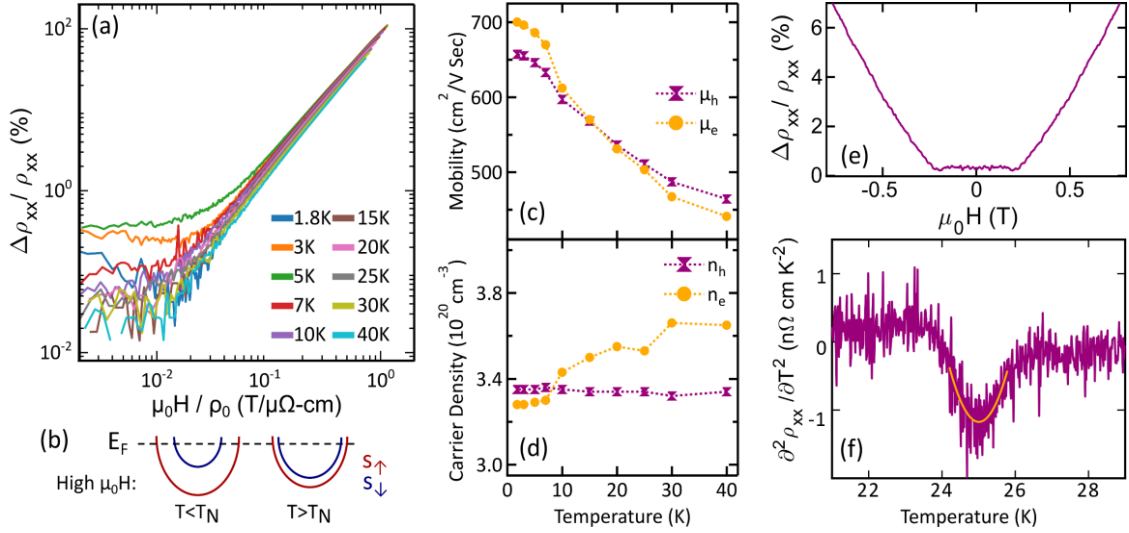

**Figure 7**: (a) Kohler scaling of magnetoresistance. (b) Schematic of magnetic field-induced exchange interactions in the electron pocket. Temperature dependencies of the mobilities (c) of charge carriers and their concentration (d) extracted from a two-band fit as described in the main text. (e) Low magnetic field magnetoresistance at 1.8 K shows the spin-flop transition. (f) Néel temperature extracted from a parabolic fit to the second derivative of the resistivity at B=0 T.

Hall coefficients measured for bulk GdSb crystals have shown a transition from *p*- to *n*-type behavior near 5K [39]. In contrast, in the 10 and 4 nm thick films a consistent *n*-type behavior is measured in **Figure 6(d)** and **Figure S7(c)**, indicating that either a carrier density mismatch or reduced hole mobility leads to a net negative Hall coefficient. Given the temperature dependence of the Hall coefficient at low temperatures <10K, the higher electron mobility likely leads to the *n*-type behavior.

In **Figure 7(a),** the Kohler plot [64] shows the effect of magnetic scattering, mobility fluctuations, and electron-electron interactions on varying relaxation times. At low fields, the magnetoresistance behavior can be separated into two temperature regimes: low temperatures ($T \ll T_N$) and high temperatures ($T \sim T_N$, $T > T_N$). At low temperatures, spin-scattering due to the spin-flop transition and quantum interference effects leads to a slower, non-parabolic rise in magnetoresistance observed up to ~2 T (~0.15 T/μΩ-cm). At high temperatures, quantum interference is suppressed, and spin scattering persists for all fields in the paramagnetic phase and results in a magnetoresistance scaling of $MR \propto (\mu_0 H/\rho_0)^{1.58}$, deviating from the ideal value of 2 due to potential carrier concentration and mobility imbalance.

High magnetic fields are expected to lead to strong exchange splitting at the electron pocket below $T_N$ (see **Figure 7(b)**), potentially leading to a change in carrier compensation. However, based on DFT calculations the ferromagnetic phase of GdSb remains charge compensated (**Table 2**), indicating that no significant carrier concentration change is expected at $T < T_N$ or at high magnetic fields.

**Figure 7(c-d)** shows the temperature dependence of the charge carrier mobilities obtained from simultaneously fitting the Hall and longitudinal magnetoresistance with a 2-band model. The hole and electron carrier density extracted from multiband fits to the magnetotransport data lies between the concentrations extracted from ARPES fits and SdH oscillations, showing an electron-rich carrier density at high temperatures and a nearly compensated carrier concentration below 10 K. This spread in the calculated carrier ratios compared to the other approaches reflects the limited accuracy a simple multicarrier model has in estimating carrier concentrations compared to the more direct methods employed earlier for studying the thin film Fermi surface. More details on the magnetotransport multiband fit are provided in **Supp. Note 3**.

In **Figure 7(c),** the hole bands show a smaller Hall mobility than the electron pocket at low temperatures. The opposite occurs above the magnetic transition temperature, a trend also observed for bulk single crystals [39]. The higher electron mobility at low temperatures where scattering is reduced is consistent with observations of a smaller effective mass for the minor axis of the electron pocket vs. the hole bands as measured by ARPES measurements (**Table 1**). After crossing $T_N$ at 25 K to lower temperatures, an accelerated increase in carrier mobilities is seen in **Figure 7(c)**, in agreement with the expected suppressed magnetic scattering at low temperatures (mainly affecting the electrons due to strong *d-f* scattering). The mobility of electrons and holes shows only a gradual linear-like increase as the temperature drops below 7 K, unlike the nearly exponential rise in mobility observed for bulk Gd-V single crystals [39,65]. The mobility saturation at low temperatures suggests that interfacial roughness and surface scattering limit carrier mobilities.

From the temperature-dependent resistivity measurements plotted in **Figure 6(a),** the kink and sharp drop in resistivity below the Néel temperature indicate that spin-disorder scattering significantly contributes to the total resistivity close to the Néel temperature. Evidence of the onset of exchange splitting in the spin-flop phase is provided in **Figure 7(e)** by the plateaued magnetoresistance at low magnetic fields, which decreases around 0.2 T and begins to follow the expected linear/quadratic-like scaling ($MR \propto (\mu_0 H)^n$ $n = 1 - 2$) after crossing the spin-flop field [66]. $T_N$ is extracted from the minima in the second derivative resistivity plots, which are expected to diverge at the Néel temperature (see example fit in **Figure 7(f)**) [44].

## IV. CONCLUSIONS

In summary, we have grown epitaxial GdSb films by MBE and studied their band structure and magnetic properties with ARPES, DFT, and magnetotransport. The growth window of high-quality single-crystal GdSb films on a lattice-matched III-V structure has been established, resulting in record magnetoresistance values for the given film thicknesses among RE-V. Our comprehensive study of GdSb films paves the way to understand the relationship between the electronic and transport properties of RE-Vs films vs. their bulk crystals and other films in semimetallic systems with similar chemistry, such as the Heusler compounds: GdXV (X=Pt, Pd, V=Sb, Bi).

We have identified magnetic and interface scattering mechanisms in GdSb, limiting the mobilities of thin RE-V films and ultimately resulting in lower magnetoresistance values than bulk single crystals and reduced magnetoresistance in thinner films. ARPES experiments of the GdSb layer showed quantization effects of energy levels in the electron and hole pockets. We did not detect these features in magnetotransport quantum oscillations due to the higher surface scattering and larger effective mass of the higher energy quantum well subband levels. Tuning the quantum level subband energy position by varying the film thickness could serve as a potential route for engineering carrier compensation and mobilities for improved magnetoresistive behavior.

While we do not see evidence of a *p-d* band inversion in ARPES measurements performed for the antiferromagnetic state, our DFT calculations show the near onset of band overlap for strong exchange splitting at low temperatures and high magnetic fields in the forced ferromagnetic phase. DFT calculations treating the *4f* electrons as core levels in a non-magnetic phase describe well the band structure of GdSb below the Néel temperature. By introducing *4f* electron AFM ordering into the valence band structure in our DFT calculations, we have obtained a more accurate description of the split-off band position and total carrier density of the electron and hole bands.

## ACKNOWLEDGEMENTS

Synthesis of thin films, development of a prototype ultra-high vacuum suitcase, ARPES experiments, and theoretical work were supported by the U.S. Department of Energy (contract no. DE-SC0014388). Development of the growth facilities and low-temperature magnetotransport measurements were supported by the Office of Naval Research through the Vannevar Bush Faculty Fellowship under award no. N00014-15-1-2845. Scanning probe studies were supported by NSF (award number DMR-1507875). This research used resources of the Advanced Light Source, which is a DOE Office of Science User Facility under contract no. DE-AC02-05CH11231. Use of the Stanford Synchrotron Radiation Lightsource, SLAC National Accelerator Laboratory, is supported by the U.S. Department of Energy, Office of Science, Office of Basic Energy Sciences under Contract No. DE-AC02-76SF00515. We acknowledge the use of shared facilities of the NSF Materials Research Science and Engineering Center (MRSEC) at the University of California Santa Barbara (DMR 1720256). DFT calculations used the National Energy Research Scientific Computing Center (NERSC), a U.S. Department of Energy Office of Science User Facility operated under contract no. DE-AC02-05CH11231. H. S. I. gratefully acknowledges support from the UC Santa Barbara NSF Quantum Foundry funded via the Q-AMASE-i program under award DMR-1906325 and support for further developments of the vacuum suitcases. D.Q.H acknowledges support from NSF through the University of Delaware Materials Research Science and Engineering Center, DMR-2011824. We thank Connor P. Dempsey, Elliot Young, and Maxwell Bocheff for technical assistance and the construction of new ultra-high vacuum experimental setups.

## APPENDIX: CARRIER DENSITY VALUES

The experimental Fermi surface volumes in **Table 1**, and **Table 3** are calculated assuming an elliptical surface for the $\alpha$ bands, and spherical surfaces for the hole pockets. The total carrier density is calculated as discussed in [67]: $n_e = 3n_\alpha$, $n_h = n_\beta + n_\delta$, $n_{\alpha/\beta/\delta} = \frac{V_F}{4\pi^3}$ and $V_F^{\beta/\delta} = \frac{4\pi}{3}r^3$ $V_F^\alpha = \frac{4\pi}{3}\left(k_F^{minor}\right)^2 k_F^{major}$ with the radius r for the hole pockets described below. The Fermi volume model accuracy was confirmed by calculating the carrier density from the DFT Fermi wavevectors in **Table 2** (marked with *). The degree of compensation between the hole and electron carriers was found to depend mainly on the approximated shape of the $\delta$ pocket Fermi volume, similar to earlier observations made for YSb [52] and other La-V compounds [27]. The non-spherical band-warping in the heavy hole $\delta$ band (and smaller distortion in the light hole $\beta$ band) is accounted for by investigating three possible Fermi surface shapes and selecting the model that best describes the DFT carrier density using the DFT Fermi wavevectors for the analytical calculation. The sphere model (b) matches best the extracted density of states in **Table 2**, and the two other models serve as lower and upper bounds for the carrier density:

a. A Fermi surface composed of two opposite pyramids sharing a base. The pyramid square base area is $A^\delta = \left(2k_F(\delta_{\bar{X}-\bar{\Gamma}-\bar{X}})\right)^2$ and the height is $h^\delta = k_F(\delta_{\bar{M}-\bar{\Gamma}-\bar{M}})$, and $n_{\delta\,(DFT)}$ =1.91 $10^{20}$ cm$^{-3}$.

b. Sphere: The radius being $r = \frac{k_F(\delta_{\bar{X}-\bar{\Gamma}-\bar{X}}) + k_F(\delta_{\bar{M}-\bar{\Gamma}-\bar{M}})}{2}$ and a total volume $V_F = \frac{4\pi}{3}r^3$, $n_{\delta\,(DFT)}$=3.62 $10^{20}$ cm$^{-3}$.

c. Average volume of two spheres. $V_F = \frac{\frac{4\pi}{3}(r_1^3 + r_2^3)}{2}$ where: $r_1 = k_F(\delta_{\bar{X}-\bar{\Gamma}-\bar{X}})$, $r_2 = k_F(\delta_{\bar{M}-\bar{\Gamma}-\bar{M}})$, and $n_{\delta\,(DFT)}$= 3.80 $10^{20}$ cm$^{-3}$.

The magnetotransport quantum oscillation carrier densities in **Table 3** were calculated from the FFT frequency of the SdH oscillations $f_{FFT}$, using the Onsager relation: $f_{FFT} = \frac{\Phi_0 A_{ext}}{2\pi^2}$. $\Phi_0$ is the magnetic flux quantum, and $A_{ext}$, the extremal orbit area for $\alpha_\perp, \beta, \delta$, is assumed to be circular: $A_{ext} = \pi k_F^2$. We extract the electron pocket semi-minor Fermi wave vector, $k_F(\alpha_{minor})$, and the hole pockets $k_F(\beta)$ and $k_F(\delta)$. The electron pocket $k_F(\alpha_{major})$ is calculated assuming an ellipsoidal band with a wavevector ratio of $\frac{k_F(\alpha_{major})}{k_F(\alpha_{minor})} = \frac{3.79}{1}$ (derived from the DFT $k_F$ ratio in the AFM phase, see **Table 2**). The Fermi surface volumes are then calculated from the SdH Fermi wavevectors as described earlier.

# Epitaxial growth, magnetoresistance, and electronic band structure of GdSb magnetic semimetal films: Supplementary Materials

Hadass S. Inbar*[1], Dai Q. Ho[2,3], Shouvik Chatterjee[4#], Mihir Pendharkar[4♦], Aaron N. Engel[1], Jason T. Dong[1], Shoaib Khalid[2†], Yu Hao Chang[1], Taozhi Guo[1‡], Alexei V. Fedorov[5], Donghui Lu[6], Makoto Hashimoto[6], Dan Read[4,7], Anderson Janotti[2], Christopher J. Palmstrøm[1,4*]

[1]Materials Department, University of California Santa Barbara, Santa Barbara, CA 93106, USA
[2]Department of Materials Science and Engineering, University of Delaware, Newark, DE 19716, USA
[3]Faculty of Natural Sciences, Quy Nhon University, Quy Nhon 55100, Vietnam
[4]Electrical and Computer Engineering Department, University of California Santa Barbara, Santa Barbara, CA 93106, USA
[5]Advanced Light Source, Lawrence Berkeley National Laboratory, Berkeley, CA 94720, USA
[6]Stanford Synchrotron Radiation Lightsource, SLAC National Accelerator Laboratory, CA, USA
[7]School of Physics and Astronomy, Cardiff University, Cardiff CF24 3AA, UK
(Dated: October 3, 2022)

## Supp. Note 1

**MBE growth of GdSb and surface characterization**

<u>MBE growth:</u> High purity Gd (4N, Materials Preparation Center, Ames Laboratory), Ga (7N, United Mineral and Chemical Corporation, UMC), Al (6N5, UMC) In (7N, UMC), and Be were evaporated from effusion cells, and Sb (7N, UMC) was supplied from a valved cracker cell as $Sb_2$. GaSb wafers (Wafer Technology Ltd.) were mounted on tungsten or tantalum substrate holders by Gallium bonding, and the temperature of the substrate was measured using a thermocouple and infrared pyrometer set to an emissivity of 0.62-0.67 (calibrated at 540°C, the thermal desorption temperature of the surface oxide under $Sb_2$ overpressure). Atomic fluxes for Al, Ga, In, and $Sb_2$ were measured with an ion-gauge (excluding Gd to avoid cathode poisoning due to the high reactivity and oxidation tendency of Gd [1]) and calibrated against RHEED intensity oscillations for Sb-rich surface reconstructions (In, Ga, Al flux) and Ga-rich surfaces (Sb flux) on GaSb (001). The GdSb RHEED intensity oscillations during growth were taken after several monolayers of GdSb film growth so as not to be affected by an embedded growth mode [2].

After the GaSb (001) native oxide desorption, a ~100 nm thick GaSb buffer layer was grown, followed by a ~1μm metamorphic buffer layer grown to match the GdSb bulk lattice constant of 6.219Å [3]: $In_{0.25}Al_{0.75}Sb$ for transport measurements (referred to as InAlSb), and Be-doped $In_{0.32}Ga_{0.68}Sb$ (referred to as InGaSb) for photoemission studies. The buffer layer was nucleated at 340-350°C and grown <370°C at a rate of ~2.2 Å /sec. **Figure S1(a-f)** shows the evolution of the RHEED pattern during GdSb (001) epitaxial growth. An Sb-rich C(2×6)/(1×3) surface reconstruction is present for all III-V layers, which evolves into a (1×1) un-reconstructed surface diffraction pattern upon GdSb growth. The GdSb growth rate was 0.03 unit cells/sec (0.187Å/sec), yielding weak RHEED intensity oscillations (**Figure S1(g)**). The elemental Gd flux was also determined *ex-situ* from the finite thickness fringes measured by x-ray diffraction (XRD). Stoichiometric growth conditions resulted in the highest quality GdSb, especially when grown directly (strained) on GaSb.

Maintaining the narrow growth window of GdSb (see **Figure S1(h)**) becomes challenging for thick film growth (>30 nm) due to unintentional radiative heating from the Gd effusion cell changing the substrate temperature. Growths with high $Sb_2$ fluxes or too low temperatures (<420°C) led to a significant amount of solid Sb adsorbed on the surface resulting in a hazy polycrystalline ring observed in RHEED, whereas too low of an $Sb_2$ flux or high-temperature growths led to Ga and In adatoms to accumulate on the surface and react with excess Gd, determined from 3D diffraction patterns seen in RHEED. A thin interlayer (4 monolayers) of AlSb was added and served as a barrier for Ga/In surface diffusion from the buffer layer, allowing higher growth temperatures and lower $Sb_2$/Gd flux ratios during GdSb growth.

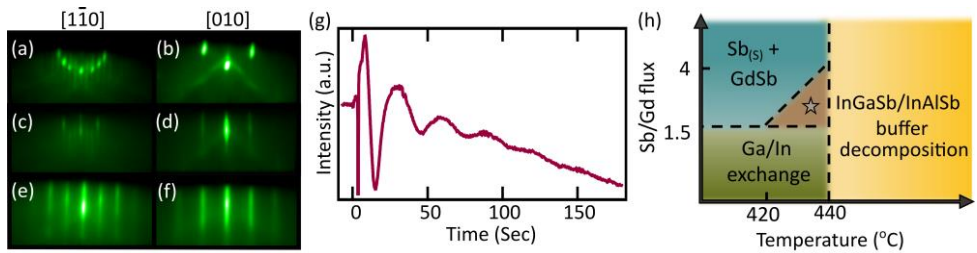

**Figure S1:** RHEED patterns measured along the [1-10] and [010] azimuths for the (a-b) GaSb substrate, (c-d) InAlSb metamorphic buffer layer, and (e-f) GdSb film. (g) GdSb RHEED intensity oscillations and (h) a schematic of the experimental phase diagram determined showing ideal GdSb growth conditions marked by a star.

<u>X-ray and UV photoemission spectroscopy of GdSb:</u> The X-ray photoemission spectroscopy (XPS) measurements of a 30 nm thick GdSb film in **Figure 2(a-c)** were performed at UC, Santa Barbara using monochromic Al k$\alpha_1$ as the X-ray source (1486.6 eV) at an emission angle of 55° in a Surface Science Laboratories SSX-100 ESCA system. Our UV-photoemission scan agrees with the predicted deep-lying energy position of the occupied 4f levels in our DFT calculations, ~8.2eV below the Fermi level.

Additional details on thin-film characterization: The film surface morphology was investigated with Nomarski optical microscopy and atomic force microscopy. However, since the final island density in the GdSb stack is primarily determined by the stabilization of the InAlSb layer at high deposition temperatures, a detailed investigation into InAlSb growth and temperature-time-$Sb_2$ flux decomposition conditions is required to establish the exact role the AlSb interlayer has in controlling In/Ga surface diffusion and RE-III-V interface reactions.

Structural quality, crystal order, and film thickness were monitored with triple-axis XRD and X-ray reflectivity (XRR) oscillations. Low-temperature magnetotransport measurements were carried out for 10 nm and 4 nm thick films in a physical property measurement system (PPMS-14T; Quantum Design) using a Van der Pauw geometry with annealed Indium electrical contacts, with $R_{xx}$ measured along [110]. Magnetotransport curves are symmetrized for $R_{xx}$ magnetoresistance and antisymmetrized for the Hall effect ($R_{xy}$).

## Supp. Note 2

**ARPES and DFT calculations of Fermi wavevectors and band positions in GdSb**

ARPES data collection and analysis: ARPES measurements at ALS were conducted at 11K and were acquired with a Scienta R4000 hemispherical analyzer with linear horizontal light. ARPES measurements at SLAC were done for Sb capped films prepared by thermally desorbing ~1μm of the Sb cap layer at 430°C for at least 1 hour. The SLAC measurements were conducted at a temperature of 20K, obtained using circularly polarized light, and acquired by a Scienta Omicron DA30L hemispherical analyzer. An inner potential of 12eV is determined, similar to previous ARPES studies of RE-Vs [6–8]. 20 nm and 4 nm thick films were studied via ARPES.

DFT calculations - additional details: Kohn-Sham orbitals in DFT [13,14] were expanded using a plane-wave basis set with a 400 eV energy cutoff. Interactions between ion cores and valence electrons were described by the projector augmented wave (PAW) method [15]. For the magnetic phase calculations, we used a rhombohedral unit cell consisting of 4 atoms and the primitive cell of an FCC crystal structure with 2 atoms to simulate the AFM and FM states, respectively. An $8 \times 8 \times 8$ $\Gamma$-centered k-point mesh was used for integration over the first Brillouin zone. The band structure was unfolded for the AFM calculation to directly compare with ARPES data [16,17]. The configurations of valence shells of Gd and Sb are $4f^75s^25p^65d^26s^1$ and $5s^25p^3$, respectively. Spin-orbit coupling was included self-consistently in all calculations. The calculated Fermi wavevectors, band extrema, and carrier densities for the AFM, FM, and non-magnetic phases are reported in **Table 2**. In **Table S1**, we present the calculated carrier ratio and its dependence on the relative position of the Fermi level. The charge carrier concentrations were determined using the SKEAF code [18], which uses the Fermi volume obtained from the Wannier90 program with the Brillouin zone sampling consisting of 1,000,000 k-points [19]. The carrier densities from the non-magnetic phase were used to estimate the carrier concentrations and degree of compensation in the AFM phase due to the similar Fermi wavevectors obtained (**Table 2**) and difficulty in calculating the AFM carrier density directly from the DFT density of states.

**Table S1:** Effect of Fermi level position on electron-hole compensation, obtained from DFT calculations for the non-magnetic phase and assuming a rigid band structure.

| $E_F$ (eV) | -50 meV | 0 | +50 meV |
|---|---|---|---|
| $n_e$ ($10^{20} cm^{-3}$) | 3.289 | 4.198 | 5.234 |
| $n_h$ ($10^{20} cm^{-3}$) | 4.863 | 4.192 | 3.596 |
| $n_e/n_h$ | 0.676 | 1.001 | 1.456 |

Band extrema and effective masses in **Table 1**: To account for the non-parabolic dispersion of the electron pocket, the conduction band is fitted to a hyperbolic function along the semimajor and semiminor ($\alpha_{minor}$, $\alpha_{major}$) axes [9]. A quartic polynomial fit to the hole δ pocket was used to assess the valence band maximum binding energy above the Fermi level, and linear fits near the Fermi level were used for extracting the Fermi wavevectors and effective masses of the hole δ and β pockets. Effective masses for all carriers are determined from the first derivative near $E_F$: $m^* = \hbar^2 k_F \left(\frac{dE(k)}{dk}\right)^{-1}$. Two hole bands (β, δ) cross the Fermi level with an extrapolated peak at 0.31eV and the spin-orbit coupled splitting at the Γ point calculated from ARPES experiments: $E_{\delta/\beta}(\Gamma)-E_\gamma(\Gamma)$=0.52 eV is consistent with other experimental values extracted for CeSb [8] yet smaller than DFT predictions in **Table 2** of 0.86 eV.

## Supp. Note 3

**Exchange spin-splitting, SdH oscillations, and magnetotransport models**

Exchange Spin-Splitting Under Magnetic Fields: Due to the large distance of the *4f* electrons from the Fermi level, the occupied *4f* levels have weak coupling to the valence band *p*-orbitals, whereas significant *d-f* Coulomb exchange interaction results with exchange splitting of the electron conduction band. In the spin-flop phase, the exchange splitting of the electron pocket band should increase linearly with magnetic field due to spin reorientation. The magnetization in GdSb was shown to linearly increase up to the critical field of 34.5 T after which GdSb enters a forced ferromagnetic phase [20]. In **Figure S2** we treat the *4f* electrons as valence electrons in a forced ferromagnetic state and calculate $E_{ex}$=483 meV as the maximum energy separation in the electron pocket expected due to exchange splitting. To evaluate the electron carrier concentration trend predicted from the 2-band fits, the SdH frequency of the electron pocket was studied as a function of temperature from 1.8 K to 15 K and the field range selected for FFT. However, throughout the field and temperature ranges selected the electron pocket FFT frequency did not shift. The absence of spin-splitting is surprising, yet can be explained by the cancellation of the SdH frequency shifts due to the nearly exact linear magnetic field dependence of the magnetization in the spin-flop phase where the expected frequency shift scales as: $\Delta F = F_{minority} - F_{majority} \propto \left(M - \frac{\partial M}{\partial B} B\right)$ [21,22]. Studies of GdBi bulk single crystals (that present a similar magnetic phase diagram and Fermi surface) also do not have any spin splitting in the SdH FFT spectrum [23]. Based on the critical field of 34.5T for GdSb [20] and the DFT calculated 483 meV exchange energy, at 14T the expected offset between minority and majority spin bands is at least 200 meV (that is before including any Zeeman splitting which should add another contribution to the gap spin-split states, albeit a smaller one assuming a g-factor of ~10 [24]).

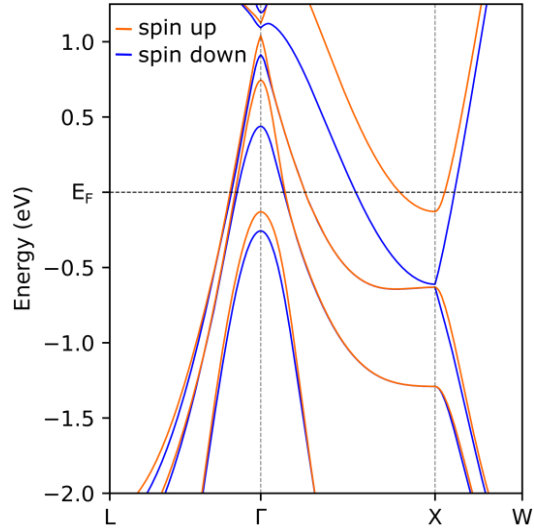

**Figure S2**: HSE06 calculated band structure of GdSb in the forced ferromagnetic state with *f* electrons treated as valence electrons. Ferromagnetic spin-exchange splitting energies for the hole and electron pockets are shown in the spin majority and minority bands.

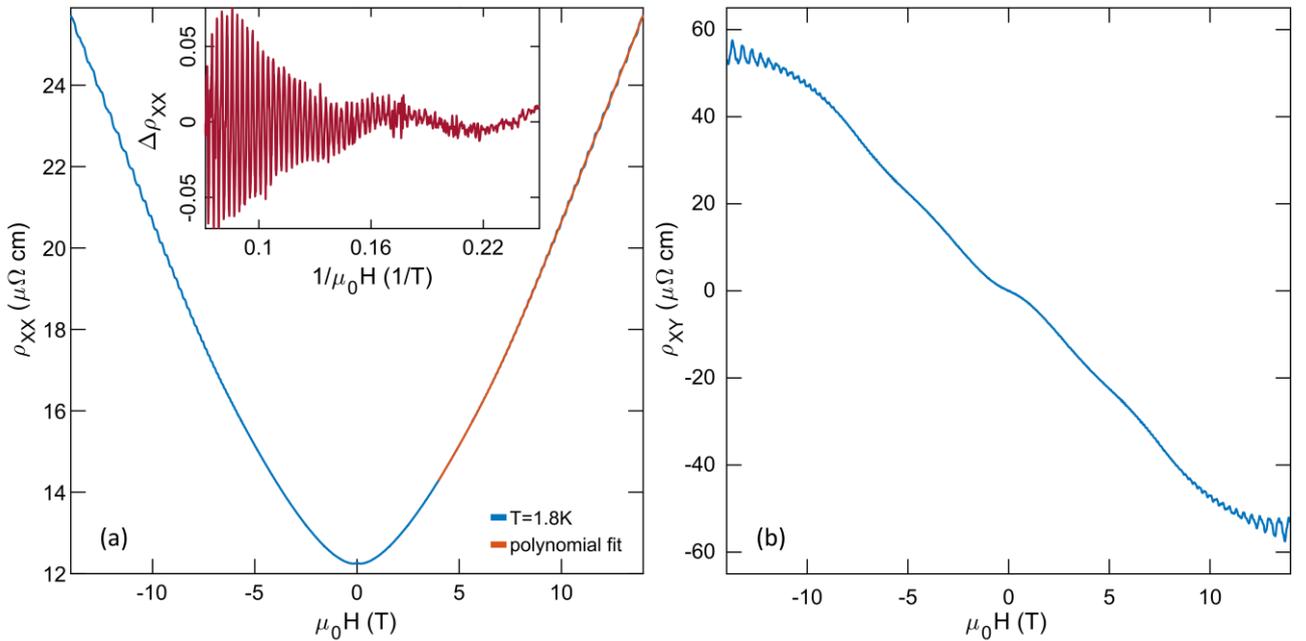

**Figure S3:** (a) Longitudinal resistivity and (b) Hall resistivity at 1.8 K. (a) 5$^{th}$ order polynomial fit used for subtracting the resistivity background. Quantum oscillations vs. inverse applied magnetic field are shown in the inset.

**SdH oscillations analysis:** In **Figure S3** the longitudinal and hall resistivity show the SdH oscillations clearly at high magnetic fields. Additional frequencies corresponding to the quantum well subband levels seen in the ARPES measurements were not observed in the quantum oscillations. Subband levels above the ground level (n=0) have a higher probability of lying near the edge of the well and therefore are more affected by interface scattering and would have lower quantum mobilities.

Multicarrier fits: In **Figure S4**, simultaneous fits to the longitudinal resistivity: $\rho_{xx} = \frac{\sigma_{xx}^{tot}}{(\sigma_{xx}^{tot})^2+(\sigma_{xy}^{tot})^2}$, and the Hall resistivity: $\rho_{xy} = \frac{\sigma_{xy}^{tot}}{(\sigma_{xx}^{tot})^2+(\sigma_{xy}^{tot})^2}$, (with $\sigma_{xx}^{tot} = \sigma_{xx}^h + \sigma_{xx}^e$ ; $\sigma_{xy}^{tot} = \sigma_{xy}^h + \sigma_{xy}^e$ and $\sigma_{xx}^i = \frac{qn_i\mu_i}{1+(\mu_i B)^2}$ ; $\sigma_{xy}^i = \frac{qn_i\mu_i^2 B}{1+(\mu_i B)^2}$ ) are performed to find the carrier density $n_i$ and mobility $\mu_i$ of a given carrier with an elementary charge $q$. While accounting for more hole bands and adding a mass anisotropy term in the electron pocket mobility values theoretically describes the Fermi surface better, adding these terms did not significantly improve our fits. Therefore, we have kept the number of fitting parameters to a minimum of 4 to avoid overfitting. The longitudinal magnetoresistance fit range was selected at relatively high fields where all magnetoresistance plots have the same scattering mechanism (see Kohler plot in **Figure 5(a)**) to avoid any contributions from spin-scattering at low fields [24] as well as any other potential change in scattering mechanism at low fields which is not accounted for in the two-band model [25].

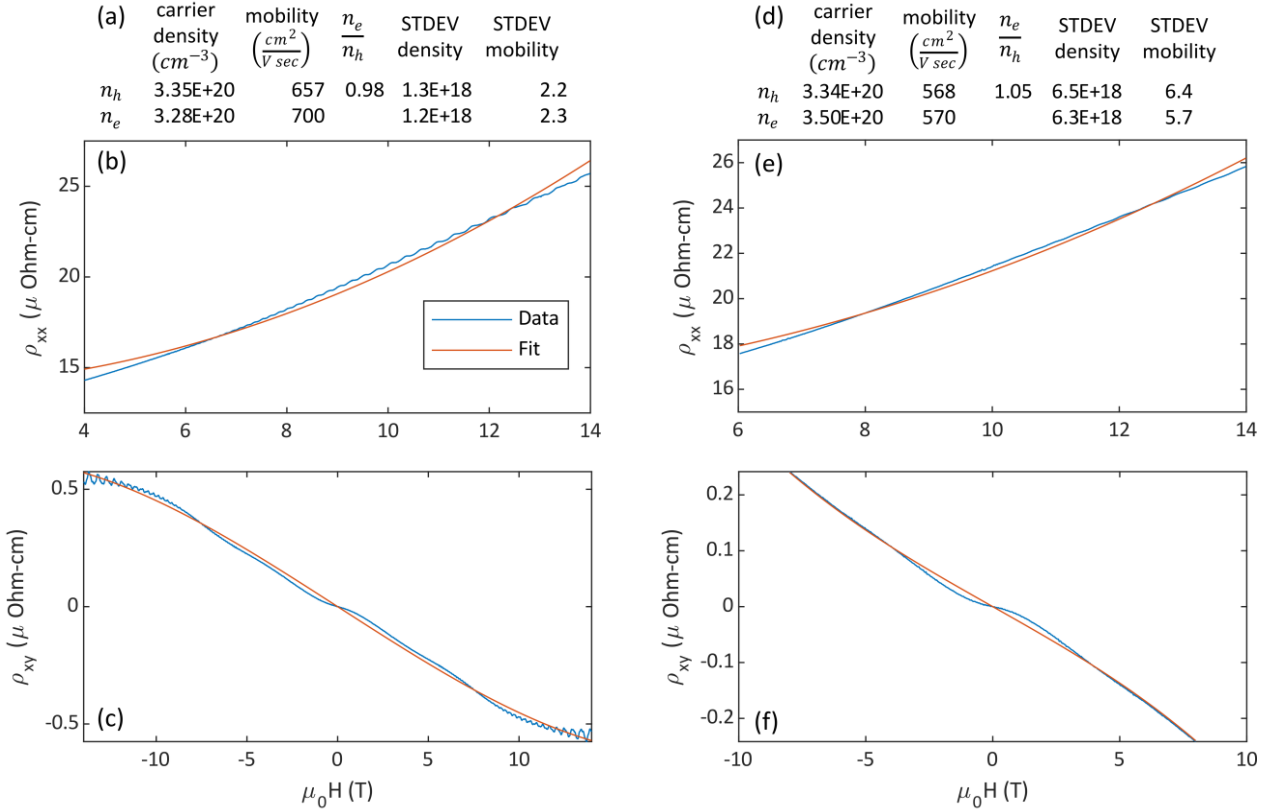

**Figure S4:** Examples of the multicarrier fits results and plots of the longitudinal resistivity ($\rho_{xx}$) and the Hall resistivity ($\rho_{xy}$) measured at 2 K (a-c) and 15K (d-f) for a 10 nm thick GdSb film.

## Supp. Note 4

**Magnetotransport in thinner films**

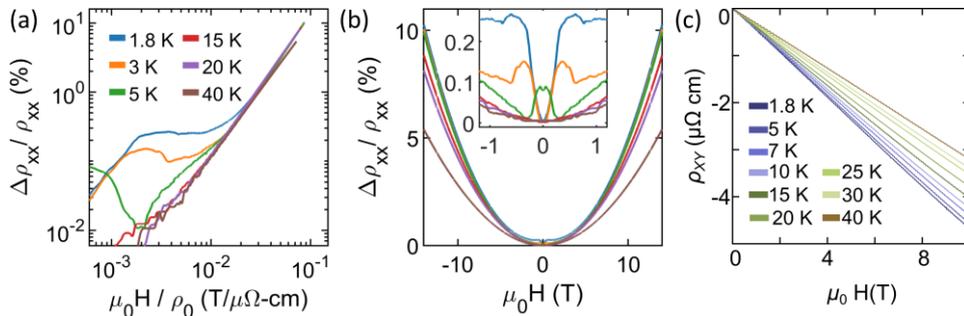

**Figure S5:** 4 nm thick GdSb film magnetoresistance (MR) at temperatures below, near, and above the Néel temperature. (a) Kohler scaling and (b) MR vs. magnetic field scaling. Inset: low field MR. At low temperatures, weak antilocalization and metamagnetic transitions are observed for B < 1T. At higher temperatures and high magnetic fields, the MR scales classically as $MR \propto (\mu_0 H/\rho_0)^{1.58}$. (c) Hall resistivity showing a more electron-dominated charge carrier behavior as the temperature decreases.

SdH oscillations in the 4 nm thick GdSb film are only resolved for the electron pocket, with a similar frequency (386 T) as the 10 nm film, in contrast to predictions of the electron pocket being lifted for thinner Gd-V films [26].

Additional frequencies corresponding to the hole pockets (β, δ) and the elliptical cross-section of the electron pocket are not observed in the FFT spectrum of the 4 nm film due to the expected lower quantum mobility for these bands and higher sensitivity of quantum mobilities to small-angle scattering primarily from misfit dislocations in the underlying InAlSb buffer layer. **Figure S5(a)** shows the Kohler plot for the 4 nm thick GdSb film. A system displaying a single scattering mechanism should have all plots with a single slope. Yet in the GdSb films, we see that at high fields $MR \propto (\mu_0 H/\rho_0)^{1.58}$ for the 10 nm thick film and $MR \propto (\mu_0 H/\rho_0)^{1.96}$ for the 4nm thick film. Given the same electron carrier density, the main difference in the exponent likely stems from a combination of (i) a closer to exact carrier compensation at thicker films and (ii) interface scattering mechanisms  the 4nm thick film which end up reducing the mobilities and altering the carrier mobility ratios. Both effects lead to a larger negative Hall coefficient observed for thinner films (**Figure S5(c))**. In LuSb films, it was shown that the hole pocket size shrinks as the film thickness decreases [27], thus a change in carrier ratios is possible upon confinement.

From the minimal mobilities and carrier densities of both charge carriers, we calculate a minimum elastic scattering length of $l_e$~270 nm. Thus for all film thicknesses studied in this paper, we are in the quantum regime where the electron wave function is coherent on length scales of the thickness of the film (i.e., d <$l_e$) [29]. The thinner 4 nm films also show at low temperatures a resistivity cusp at low fields (**Figure S5(b)**, inset), a signature of the weak antilocalization effect reflecting the strong spin-orbit coupling in GdSb. These magnetotransport signatures for thin films were present in other epitaxial RE-V and high spin-orbit coupled semimetal films of similar thickness [27,28]. In addition to weak antilocalization and potential electron-electron interactions in thin films, magnetic field-induced transitions at low temperatures are also present: a spin-flop transition is seen at 0.2T in **Figure S5(b)**, inset as well as an intermediate metamagnetic phase that is stable between 0.2T to 0.5T. Changes in resistance due to weak magnetic impurity states [3] are ruled out based on the temperature dependence of the spin-flop field, and weak localization and electron-electron interactions effects are also excluded by observing similar behavior in 20 nm thick films. Jumps and kinks seen in the magnetoresistance curve at low fields near the spin-flop transition are similar to earlier observations of metamagnetic transitions in other RE-Vs such as HoBi [30] and ErAs [24].